\documentclass[twocolumn,showpacs,aps,pra]{revtex4-1}

\usepackage{amsmath,amsfonts,amssymb}
\usepackage{wrapfig}
\usepackage{graphicx}
\usepackage{bbm}
\usepackage{color}
\usepackage{url}
\usepackage{ulem}

\begin{document}

\title{Discriminating quantum correlations with networking quantum teleportation}

\author{Shih-Hsuan Chen$^{1}$}
\author{He Lu$^{2,3,4}$}
\author{Qi-Chao Sun$^{3,4}$}
\author{Qiang Zhang$^{3,4}$}
\author{Yu-Ao Chen$^{3,4}$}
\author{Che-Ming Li$^{1,3,5,6,}$}
\email{cmli@mail.ncku.edu.tw}

\affiliation{$^{1}$Department of Engineering Science, National Cheng Kung University, Tainan 70101, Taiwan}
\affiliation{$^{2}$School of Physics, Shandong University, Jinan 250100, China}
\affiliation{$^{3}$Shanghai Branch, National Laboratory for Physical Sciences at Microscale and Department of Modern Physics, University of Science and Technology of China, Shanghai 201315, China}
\affiliation{$^{4}$Synergetic Innovation Center of Quantum Information and Quantum Physics, University of Science and Technology of China, Hefei, Anhui 230026, China}
\affiliation{$^{5}$Center for Quantum Technology, Hsinchu 30013, Taiwan}
\affiliation{$^{6}$Center for Quantum Frontiers of Research \& Technology, National Cheng Kung University, Tainan 701, Taiwan}

\begin{abstract}
The Bell inequality, and its substantial experimental violation, offers a seminal paradigm for showing that the world is not in fact locally realistic. Here, going beyond the scope of Bell's inequality on physical states, we show that quantum teleportation can be used to quantitatively characterize quantum correlations of physical processes. The validity of the proposed formalism is demonstrated by considering the problem of teleportation through a linear three-node quantum network. A hierarchy is derived between the Bell nonlocality, nonbilocality, steering and nonlocality-steering hybrid correlations based on a process fidelity constraint. The proposed framework can be directly extended to reveal the nonlocality structure of teleportation through any linear many-node quantum network. The formalism provides a faithful identification of quantum teleportation and demonstrates the use of quantum-information processing as a means of quantitatively discriminating quantum correlations.
\end{abstract}

\maketitle

\section{INTRODUCTION}
\begin{figure}[t]
\includegraphics[width=8.3cm]{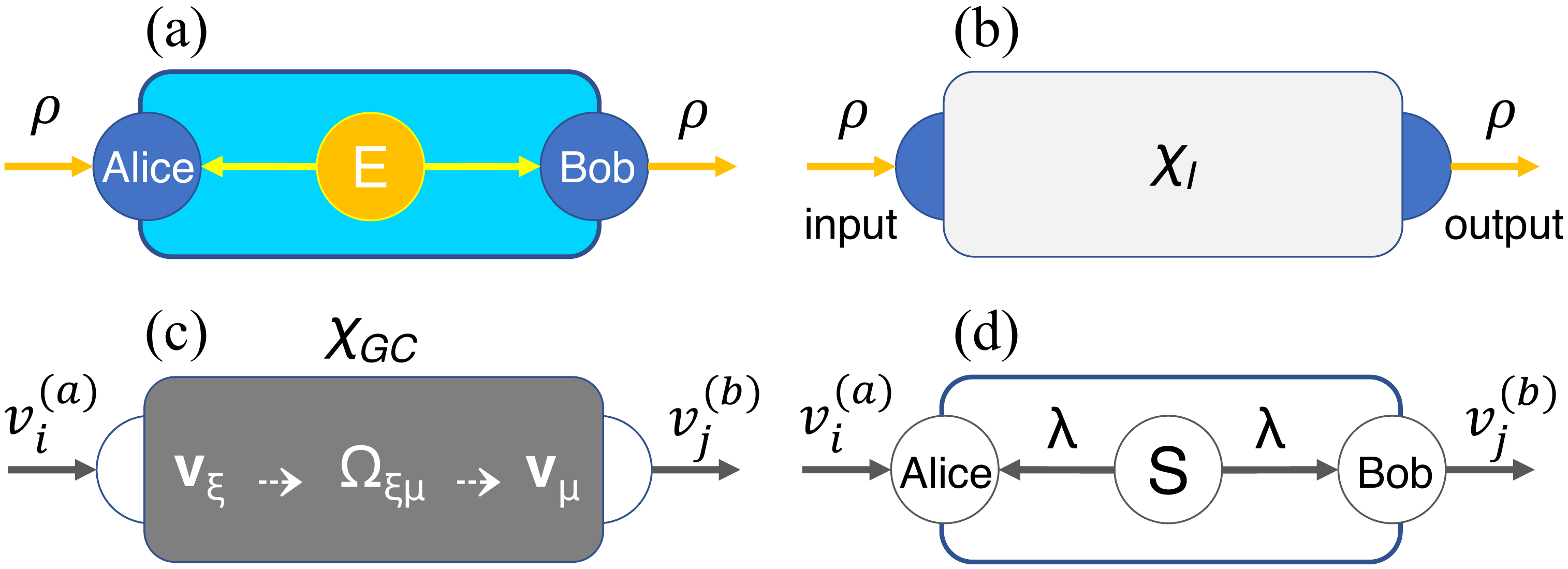}
\caption{Quantum teleportation and its classical mimicry. (a) Implementation of teleportation. The sender node (Alice) and receiver node (Bob) of the network first share a qubit pair from an entanglement source (E). Alice performs quantum joint-state measurement on the transmitted qubit with a state $\rho$ and half of the entangled pair held by her in the basis of Bell states (not shown). She then sends her measurement result to Bob. Finally, depending on Alice's measurement outcome, Bob performs local operations on his half of the entangled pair to recover the unknown state. (b) Notably, this input-output procedure acts as an identity quantum process $\chi_{I}$. (c) The proposed formalism of a genuinely classical process (GCP) $\chi_{GC}$ is used to simulate $\chi_{I}$ and derive faithful criteria for experiments. (d) The GCP formalism is sufficiently general to encompass the local hidden variable (LHV) model for mimicking teleportation using a classical source (S).}\label{basicidea}
\end{figure}

Quantum teleportation \cite{Bennett96} enables networking participants to move an unknown quantum state between the nodes of a quantum network \cite{Gisin07}. Quantum teleportation experiments have been realized in laboratories \cite{Pirandola15,Bouwmeester97,Zhang06}, free space \cite{Yin12,Xia18}, and even ground to satellite \cite{Ren17}. The ideal teleportation of a qubit with an unknown state $\rho$ acts as an identity unitary transformation, $\chi_{I}$, on the transmitted state, i.e., $\chi_{I}(\rho)=\rho$, as illustrated in Figs.~\ref{basicidea}(a) and \ref{basicidea}(b). However, if such a quantum process is attacked by an eavesdropper, or manipulated by untrusted networking participants, the performance of all the networking tasks underlying the quantum teleportation process becomes questionable \cite{Pirandola15}. Thus, the problem of identifying genuinely quantum teleportation through quantum networks, and ruling out any classical strategies of mimicry, poses an interesting but significant challenge to both quantum-information processing and practical implementation. In particular, while it is known that networking teleportation is fueled by quantum operations and entangled pairs shared between participants, it is not yet clear how the teleportation task can be utilized to quantitatively characterize the quantum correlations underlying the network.

In order to tackle this problem, we introduce the concept of a genuinely classical process (GCP) to simulate the ideal quantum teleportation process, $\chi_{I}$, and provide a strategy for mimicking teleportation by classical physics. The proposed formalism not only provides a benchmark of faithful teleportation, but also gives the means to classify the quantum correlations between the quantum nodes. In contrast to existing theories, which utilize the \textit{state characteristics} to verify teleportation \cite{He15,Chiu16,Cavalcanti17} and quantum correlations (e.g., Bell nonlocality \cite{Brunner14}, nonbilocality \cite{Branciard10,Branciard12,Saunderse17,Carvacho17}, and non-$N$ locality \cite{Tavakoli14,Rosset16,Tavakoli16,Lee18}), the proposed formalism is truly task-oriented, and is thus well suited to the characterization of general many-node networking teleportation and its underlying quantum resources. Moreover, the formalism can be readily implemented in a wide variety of present experiments on teleportation, as will be shown in subsequent sections.

\section{Genuinely classical processes}
We define a GCP as a set of three steps describing the system state and its evolution. In particular, we assume that the input particle undergoes a generic physical process and decays into a classical system that is considered to be a physical object with properties satisfying the assumption of realism \cite{Brunner14}. The system then evolves in accordance with classical stochastic theory \cite{Breuer&Petruccione02}. Finally, once the process is complete, the system resides in its final output state with properties satisfying the assumption of realism.

Since a GCP treats the initial system as a physical object with properties satisfying the assumption of realism, the system can be modeled as a state described by a fixed set of physical properties $\textbf{v}_{\xi}$. Assume that the system is described by three properties, say $V_{1}$, $V_{2}$ and $V_{3}$, where each property has two possible states. There therefore exist $2^{3}=8$ sets underlying the classical object: $\textbf{v}_{\xi}(\text{v}_{1},\text{v}_{2},\text{v}_{3})$, where $\text{v}_{1},\text{v}_{2},\text{v}_{3}\in\{+1,-1\}$ represent the possible measurement outcomes for $V_{1}$, $V_{2}$ and $V_{3}$, respectively. The subsequent classical evolution of the system changes the system from an initial state $\textbf{v}^{(a)}_{\xi}$ to a final state $\textbf{v}^{(b)}_{\mu}$ according to the transition probabilities $\Omega_{\xi\mu}$. Therefore, the relationship between a specific input state of the $i$th physical property of the system, e.g., $\text{v}_{i}=v^{(a)}_{i}$, and a specific output state of the $j$th physical property $v^{(b)}_{j}$ can be characterized by
\begin{equation}
P(v^{(b)}_{j}|v^{(a)}_{i})=\sum_{\xi,\mu}P(\textbf{v}^{(a)}_{\xi}|v^{(a)}_{i})\Omega_{\xi\mu}P(v^{(b)}_{j}|\textbf{v}^{(b)}_{\mu}).\label{resultingstate}
\end{equation}
[see Fig.~\ref{basicidea}(c)]. Let process tomography (PT), a particular application of the quantum operations formalism \cite{Chuang96,Nielsen&Chuang00}, be used to systematically exploit the experimentally measurable quantities given in Eq. (\ref{resultingstate}). After PT, the GCP can be completely characterized with a positive Hermitian matrix, called the process matrix, of the form
\begin{equation}
\chi_{GC}=  \frac{1}{4}\left[ \begin{matrix}
    \hat{I}_{gc}+\hat{V}_{gc,3} & \hat{V}_{gc,1}+i\hat{V}_{gc,2} \\
    \hat{V}_{gc,1}-i\hat{V}_{gc,2} & \hat{I}_{gc}-\hat{V}_{gc,3}
    \end{matrix}
\right],\label{process_tomography}
\end{equation}
where $\hat{I}_{gc}\equiv\rho_{v^{(a)}_{i}=+1}+\rho_{v^{(a)}_{i}=-1}$ and $\hat{V}_{gc,i}\equiv\rho_{v^{(a)}_{i}=+1}-\rho_{v^{(a)}_{i}=-1}$ for $i=1,2,3$ (see Appendix \ref{derived_process_matrix}) for details. Using Eq. (\ref{resultingstate}) and state tomography \cite{Vogel89}, the density operator of the output system conditioned on a specific initial state $v^{(a)}_{i}$ is given by
\begin{equation}
\rho_{v^{(a)}_{i}}=\frac{1}{2}(\hat{I}+\sum_{j=1}^{3}\sum_{v^{(b)}_{j}=\pm 1}v^{(b)}_{j}P(v^{(b)}_{j}|v^{(a)}_{i})\hat{V}^{(b)}_{j}),\label{rhogc}
\end{equation}
where $\hat{I}$ denotes the identity operator and the observables $\hat{V}_{j}^{(b)}$ are the quantum analogs of the physical properties $V_{j}^{(b)}$ and are complementary to each other.

Notably, the above description of a system and its evolution extends the idea of the quantum output state represented by a density operator in a classical process (CP) \cite{Hsieh17}, denoted as $\chi_{C}$. As shown in Appendix \ref{comparison}, the $\chi_{GC}$ can fully describe the $\chi_{C}$.

\section{Bell nonlocality}
Suppose that a process of interest is created and its normalized process matrix, $\chi_{\text{expt}}$, is derived from experimentally available data using the PT procedure, as described above. Suppose further that the process fidelity of $\chi_{\text{expt}}$ and $\chi_{I}$ is used to evaluate the performance of the experimental process. For a given set of observables $\{\hat{V}_{j}^{(a)},\hat{V}^{(b)}_{j}|j=1,2,3\}$, if the process fidelity satisfies
\begin{equation}
F_{\text{expt}}\equiv \text{tr}(\chi_{\text{expt}}\chi_{I})> F_{GC}\equiv\max_{\chi_{GC}} \text{tr}(\chi_{GC}\chi_{I}),\label{fidelity}
\end{equation}
then the experimental process $\chi_{\text{expt}}$ is qualified as truly nonclassical and is close to teleportation. The overriding goal of Eq. (\ref{fidelity}) is to rule out the best classical mimicry of ideal teleportation $\chi_{I}$. Such a capability of genuinely classical mimicry can be evaluated by performing the following mathematical maximization task via semidefinite programming (SDP) with MATLAB \cite{Lofberg, sdpsolver}: $\max_{\chi_{GC}}\hspace{3pt}\text{tr}(\chi_{GC}\chi_{I})$,
such that $\chi_{GC}\geq 0,\hspace{3pt}\text{tr}(\chi_{GC})=1$, $\Omega_{\xi\mu} \geq 0$ $\forall\ \xi$,$\mu$. The above constraints ensure that the GCP matrix $\chi_{GC}$ satisfies both the definitions of process fidelity and a density operator. Here the computational cost depends on the number of possible transition probabilities $\Omega_{\xi\mu}$ under evaluation, and then only relies on the dimension of the input system (see Appendix \ref{Computational_cost}).

Since the observables for the PT procedure are chosen as $\hat{V}^{(a)}_{1}=X, \hat{V}^{(a)}_{2}=Y,$ $\hat{V}^{(a)}_{3}=Z$ for the input states and $\hat{V}^{(b)}_{1}=UXU^{\dag}, \hat{V}^{(b)}_{2}=UYU^{\dag},$ $\hat{V}^{(b)}_{3}=UZU^{\dag}$ for the output states, where $U=\left|0\right\rangle\!\!\left\langle0\right|+\exp(i\pi/4)\left|1\right\rangle\!\!\left\langle1\right|$ and $X$, $Y$, and $Z$ are the Pauli matrices, the clearest distinction possible is obtained between the classical result and the quantum mechanical prediction. The closest similarity to teleportation is
\begin{equation}
F_{GC}\sim0.8536,\label{fgc}
\end{equation}
under the above measurement setting (see Appendix \ref{Clearest_distinction}). (Note that the same measurement setting is used for all the networking cases presented in the remainder of the text.) It is worth noting that, for the best classical simulation (\ref{fgc}), $\chi_{GC}$ mimics $\chi_{I}$ as a phase damping process \cite{Nielsen&Chuang00}
\begin{equation}
\chi_{GC}(\rho)=0.8536\hat{I}\rho\hat{I}^{\dag}+0.1464Z\rho Z^{\dag},\label{processmatrix8536}
\end{equation}
with noise intensity $0.1464$ (see Appendix \ref{Clearest_distinction}). The performance inspection described here relies only on the preparation of four different input states \cite{four_input} and the relevant output state tomography for PT. Therefore, existing reported experiments on teleporting qubits are sufficient for checking the teleportation performance \cite{Pirandola15,Yin12,Bouwmeester97,Ren17,Xia18,Zhang06}.

It is noted that the criterion proposed above is stricter than the existing criterion used to identify faithful teleportation as a means of ruling out the measure-prepare strategy (a direct classical mimicry strategy) \cite{Pirandola15,Measure-prepare_strategy,Massar95}. The best capability for the measure-prepare strategy to mimic teleportation is $F_{\text{expt}}=0.5$, i.e., the average state fidelity of the input and output states, $\bar{F}_{\text{expt,s}}=2/3\sim0.6667$, where $\bar{F}_{\text{expt,s}}=(2F_{\text{expt}}+1)/3$ \cite{Gilchrist05}. However, according to the fidelity criterion proposed in Eqs. (\ref{fidelity}) and (\ref{fgc}), the average state fidelity is $\bar{F}_{\text{expt,s}}>\bar{F}_{GC,s}\sim0.9024$. This result implies that not all entangled states can demonstrate a teleportation process that goes beyond $\chi_{GC}$ \cite{Cavalcanti17}.

A GCP can be treated as an input-output transformation implemented by sharing local hidden variables (LHVs) between the parties involved (Alice and Bob). Equation~(\ref{resultingstate}) in the LHV model thus becomes $P(v^{(b)}_{j}|v^{(a)}_{i})=2\sum_{\lambda}P(v^{(a)}_{i}|\lambda)P(\lambda)P(v^{(b)}_{j}|\lambda)$. See Appendix \ref{process_LHV} for details. As shown in Fig.~\ref{basicidea}(d), LHV $\lambda$ describes the connection between the input state $v^{(a)}_{i}$ and the output state $v^{(b)}_{j}$. Moreover, the distribution $P(\lambda)$ determines the process matrix $\chi_{GC}$, by which one can consider the correlation of $\chi_{GC}$ as \textit{Bell local}. Therefore, since $F_{\text{expt}}>F_{GC}$, the resulting process possesses \textit{Bell nonlocality} to enable the networking participants to perform qualified experimental teleportation. This approach to testing the Bell local model is different from that of existing Bell tests, which all use Bell-like inequalities \cite{Brunner14,Branciard10,Branciard12,Saunderse17,Carvacho17,Tavakoli14,Rosset16,Tavakoli16}. (Notably, the manner in which a concrete information task together with its implementation can be described by a LHV model for networking process is not included in these standard Bell tests.)

\section{Quantum correlations of three-node network}
\begin{figure}[t]
\includegraphics[width=6.5cm]{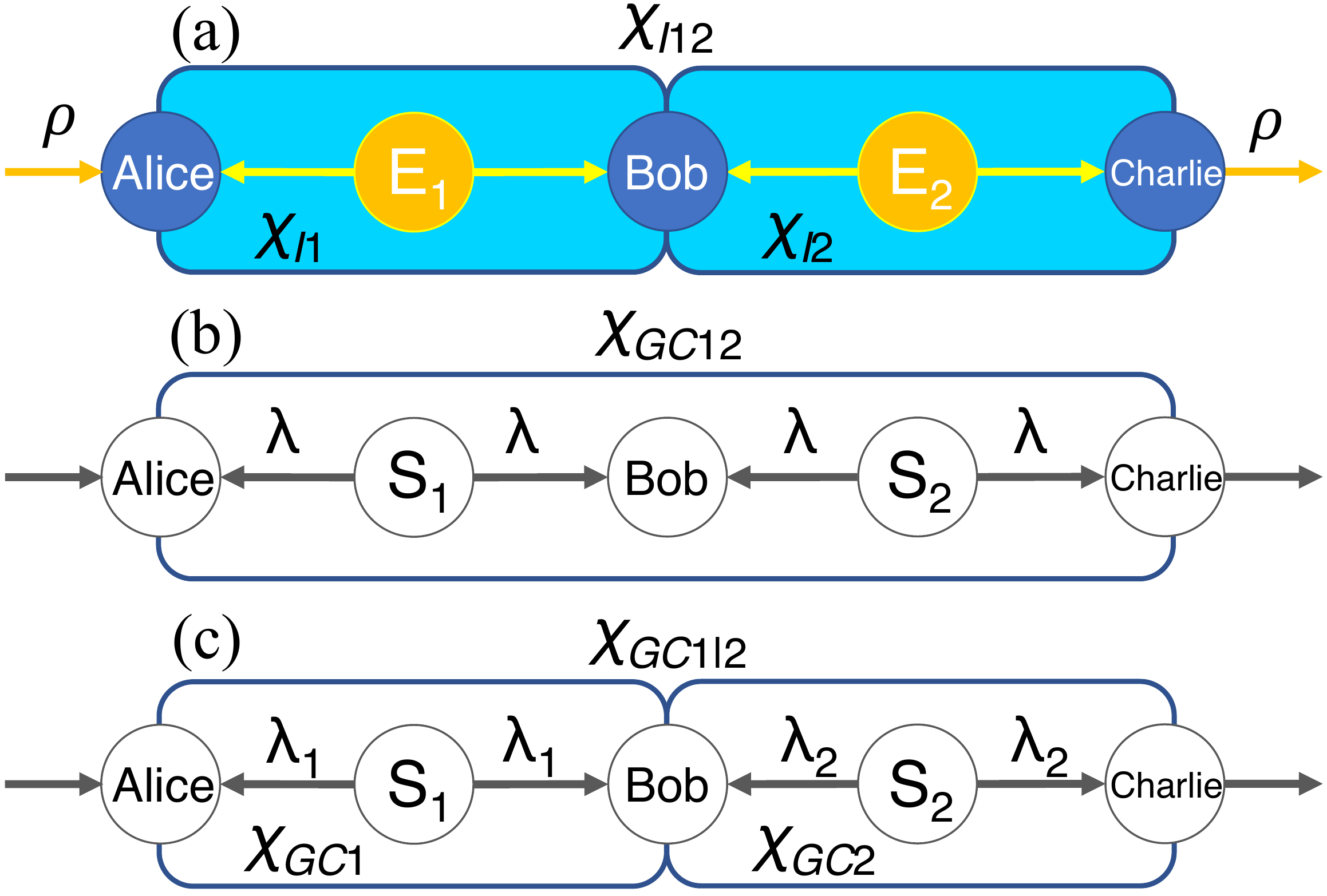}
\caption{Networking teleportation and its classical mimicry. (a) A three-node teleportation process is implemented with two entanglement sources ($E_{1,2}$) and two Bell state measurements on Alice's node and Bob's node, respectively (not shown). Two dependent classical sources ($S_{1,2}$) are used for teleportation simulation under the Bell local model, and hence (b) the resulting process $\chi_{GC12}$ is a GCP involving all three participants. By contrast, if the sources are independent, the underlying correlations become bilocal (c), and the resulting process $\chi_{GC1|2}$ is composed of two individual GCPs, i.e., $\chi_{GC1}$ and $\chi_{GC2}$.}\label{three-node}
\end{figure}

Teleporting unknown qubits through a linear network composed of three quantum nodes can be implemented by repeating the bipartite teleportation procedure twice in parallel [Fig.~\ref{three-node}(a)]. For example, assume that it is desired to teleport a qubit with state $\rho$ from the first node in a network (say, Alice) to the end node in the network (say, Charlie) through an intermediate node (say, Bob). Since the overall teleportation procedure consists of two ideal input-output subprocesses connecting Alice and Bob, $\chi_{I1}$, and Bob and Charlie, $\chi_{I2}$, respectively, the resultant process, $\chi_{I12}=\chi_{I1}\circ\chi_{I2}$, is still an identity unitary transformation with the mapping $\chi_{I12}(\rho)=\rho$, where $\circ$ denotes the concatenation operator. In other words, the general criterion given in Eq. (\ref{fidelity}) for identifying teleportation between two nodes still holds for three-node quantum networks.

From a classical viewpoint, the three-node networking task described above can be simulated using the same LHV model as that used for the two-node case. That is, the distribution of LHV $P(\lambda)$ determines the resultant GCP, where LHV $\lambda$ correlates Alice's inputs and Charlie's outputs  and then results in a specific process [Fig.~\ref{three-node}(b)]. For the measurement setting given above, the closest similarity to the three-node teleportation process that can be achieved by $\chi_{GC12}$ is quantified as
\begin{equation}
F_{GC12}\equiv \max_{\chi_{GC12}}\hspace{3pt}\text{tr}(\chi_{GC12}\chi_{I12})\simeq0.8536.\label{FGC12}
\end{equation}
The correlation of an experimental three-node qubit transmission process is then Bell nonlocal if the experimental process $\chi_{\text{expt}12}$ satisfies the criterion $F_{\text{expt}12}\equiv\text{tr}(\chi_{\text{expt}12}\chi_{I12})>F_{GC12}$.


Ideal three-node teleportation requires two entangled pairs. When transmitting qubits between distant nodes in a general network, these pairs are inevitably generated by two spatially separated independent sources \cite{Saunderse17,Carvacho17}. As a result, it is reasonable to infer that the classical strategy using a single LHV $\lambda$ to mimic teleportation can be modified.

Given the assumption of independent sources, one can reasonably assign an individual LHV to each subprocess. Let $\lambda_{1}$ and $\lambda_{2}$ be the LHVs assigned to the state transmissions between Alice and Bob, and Bob and Charlie, respectively [see Fig.~\ref{three-node}(c)]. The relation between the inputs and outputs of each subprocess is totally determined by the underlying LHVs $\lambda_{k}$ ($k=1,2$). The distribution of the LHV $P(\lambda)$ in the original LHV model for two-node state transmission thus becomes a product of the joint probability of these LHVs, $P(\lambda_{1})P(\lambda_{2})$, in the three-node case, which implies that $P(v^{(b)}_{j}|v^{(a)}_{i})=P(v^{(b)}_{j})$. Therefore, $P(\lambda_{1})$ and $P(\lambda_{2})$ determine their individual GCPs, say $\chi_{GC1}$ and $\chi_{GC2}$, respectively. The resulting GCP in the three-node network is specified by $\chi_{GC1|2}\equiv\chi_{GC1}\circ\chi_{GC2}$, where the correlation of $\chi_{GC1|2}$ is referred to as \textit{bilocal}.

The maximum fidelity of $\chi_{GC1|2}$ and $\chi_{I12}$ can be regarded as a threshold for the bilocal model. For the present case, the fidelity threshold is given as
\begin{equation}
F_{GC1|2}\equiv \max_{\chi_{GC1|2}}\hspace{3pt}\text{tr}(\chi_{GC1|2}\chi_{I12})\simeq0.7500\label{FGC1|2}
\end{equation}
(see Appendix \ref{derived_fidelities}). When each subprocess matrix is experimentally measured by PT as $\chi_{\text{expt}k}$ for $k=1,2$, the correlation of the joint process $\chi_{\text{expt}1|2}=\chi_{\text{expt}1}\circ\chi_{\text{expt}2}$ is \textit{nonbilocal} (or possesses \textit{nonbilocality}), if $F_{\text{expt}1|2}\equiv\text{tr}(\chi_{\text{expt}1|2}\chi_{I12})>F_{GC1|2}$.

If the receiver, Charlie, manipulates the received system under quantum operations and trusts his measurement equipment, then the above-mentioned bilocal model becomes a LHV-LHS (local hidden state \cite{Wiseman07}) hybrid model provided that the sources are independent. The transmission between Bob and Charlie can then be described by a classical process matrix \cite{Hsieh17}, $\chi_{C2}$, while the subprocess for Alice and Bob is specified by $\chi_{GC1}$. However, if the two subprocesses share the same pair (i.e., the sources are dependent), then the LHS model specifies the resulting process as being classical by $\chi_{C12}$. For the present measurement setting, the closest similarities between $\chi_{C12}$ and $\chi_{I12}$, and between the hybrid process $\chi_{GC1|C2}\equiv\chi_{GC1}\circ\chi_{C2}$ and an ideal teleportation process, are given as follows:
\begin{eqnarray}
&&F_{C12}\equiv \max_{\chi_{C12}}\hspace{3pt}\text{tr}(\chi_{C12}\chi_{I12})\simeq0.6830,\label{FC12}\\
&&F_{GC1|C2}\equiv \max_{\chi_{GC1|C2}}\hspace{3pt}\text{tr}(\chi_{GC1|C2}\chi_{I12})\simeq0.5985\label{FGC1|C2}
\end{eqnarray}
(see Appendix \ref{derived_fidelities}). Thus, $F_{\text{expt}12}>F_{C12}$ implies steering in $\chi_{\text{expt}12}$, while $F_{\text{expt}1|2}>F_{GC1|C2}$ implies that the networking process $\chi_{\text{expt}1|2}$ has a nonlocality-steering hybrid correlation.

\begin{figure}[t]
\includegraphics[width=8.7cm]{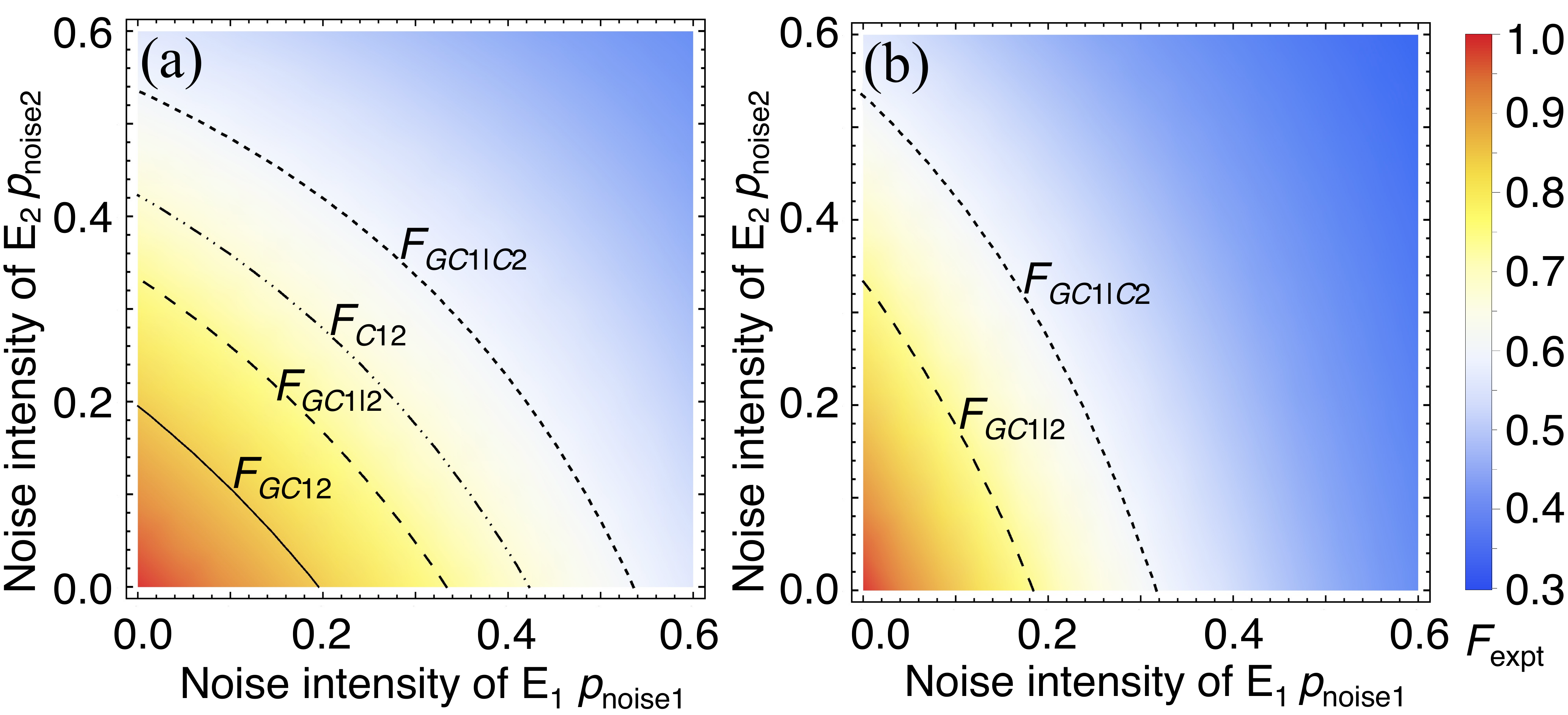}
\caption{Quantum correlations in noisy three-node teleportation. When the entangled state $\left|\phi^+\right\rangle=(\left|00\right\rangle+\left|11\right\rangle)/\sqrt{2}$ created by source $E_{k}$ mixes with white noise and becomes $\rho_{E_{k}}=(1-p_{\text{noise}k})\left|\phi^+\right\rangle\!\!\left\langle\phi^+\right|+p_{\text{noise}k}\hat{I}/4$ for $k=1,2$, the experimental process fidelities, i.e., (a) $F_{\text{expt12}},F_{\text{expt}1|2}$ and (b) $F_{\text{expt112}}$, decrease with an increasing noise intensity $p_{\text{noise}k}$. (a) Applying the fidelity thresholds given in Eqs. (\ref{FGC12})-(\ref{FGC1|C2}), the underlying quantum correlations can be discriminated in accordance with Eq.~(\ref{levels}). Meanwhile, (b) the nonbilocality and nonlocality-steering hybrid correlations can be identified through the criteria given in Eq. (\ref{levels2}).}\label{hybridprocess_noise}
\end{figure}

The fidelity thresholds in Eqs.~(\ref{FGC12})-(\ref{FGC1|C2}) suggest the existence of the following hierarchy between the Bell nonlocality, nonbilocality, steering and nonlocality-steering hybrid correlations of the teleportation process, i.e.,
\begin{equation}
\begin{split}
&F_{GC12}<F_{\text{expt12}} \ \leq 1 \ \ \ \ \ \ \  \ \ \ \text{Bell nonlocality},\\
&F_{GC1|2}<F_{\text{expt}1|2} \leq F_{GC12} \ \ \ \text{nonbilocality}, \\
&F_{C12}<F_{\text{expt12}} \leq \ F_{GC1|2}\ \ \ \ \ \text{steering},\\
&F_{GC1|C2}<F_{\text{expt}1|2} \leq F_{C12}\ \ \ \text{nonlocality steering}.
\end{split}\label{levels}
\end{equation}
Compared with the process $\chi_{GC1|2}$ under the bilocal model, the process $\chi_{GC12}$ under the Bell local model achieves a better simulation of ideal teleportation in terms of the process fidelity. Thus, the correlations of a networking process $\chi_{\text{expt}12}$ that are identified as nonlocal through (\ref{FGC12}) can always go beyond the bilocal description $\chi_{GC1|2}$. However, nonbilocality of $\chi_{\text{expt}1|2}$ does not necessarily imply the existence of Bell nonlocality. (Note that the nonbilocality, steering and nonlocality-steering hybrid correlations can be compared and analyzed in an analogous manner.)

In addition to the criteria for discriminating correlations given in Eq.~(\ref{levels}), the experimental process matrices $\chi_{\text{expt}12}$ and $\chi_{\text{expt}1}$ can be sufficient to verify the underlying correlations for teleportation. For example, consider the process fidelity $F_{\text{expt112}}\equiv\text{tr}(\chi_{\text{expt}1}\circ\chi_{\text{expt}12}\chi_{I12})$, and posit that the nonbilocality and nonlocality-steering hybrid correlations of $\chi_{\text{expt}12}$ can be identified according to the following criteria:
\begin{equation}
\begin{split}
&F_{\text{expt112}}>F_{GC1|2}\  \ \ \ \ \text{nonbilocality}, \\
&F_{\text{expt112}}>F_{GC1|C2}\ \ \ \text{nonlocality steering}.
\end{split}\label{levels2}
\end{equation}
When the correlation is bilocal, the fidelity $F_{\text{expt112}}=F_{GC1|2}$ becomes maximal when $\chi_{\text{expt}1}\circ\chi_{\text{expt}12}=\chi_{GC1|2}$. If $\chi_{\text{expt}1}\circ\chi_{\text{expt}12}=\chi_{GC1|C2}$, the fidelity $F_{\text{expt112}}=F_{GC1|C2}$ is maximum under the LHV-LHS hybrid assumption. See Fig.~\ref{hybridprocess_noise} for an illustrative example of the correlation discrimination in (\ref{levels}) and identification criteria in (\ref{levels2}) for noisy entanglement sources.

It is worth emphasizing here that $\chi_{\text{expt}1}$, $\chi_{\text{expt}2}$, and $\chi_{\text{expt}12}$ provide a complete description of what operations and errors are involved in the three-node experiment. Thus, the resulting process fidelities can be used to reveal the experimental performance by referring to the correlation hierarchy in (\ref{levels}) and identification criteria in (\ref{levels2}). Such an inspection is beneficial in evaluating and improving primitive operations in networking teleportation from the viewpoints of trusted-untrusted measurement devices and dependent-independent sources in the experiment.

\section{Non-$N$ locality}
The correlation discrimination method introduced above can be readily extended to explore the quantum correlations in general many-node teleportation networks. For example, in the following, we demonstrate the quantitative characterization of non-$N$-local correlations \cite{Tavakoli14,Rosset16,Tavakoli16} of networking teleportation involving $N$ independent entanglement sources. (Note that the nonlocality-steering hybrid correlation can be characterized in the same way.) For ideal $(N+1)$-node teleportation, the resultant process remains an identity operation $\chi_{I1N}(\rho)=\rho$, where $\chi_{I1N}=\prod_{k=1}^{N}\chi_{Ik}=\chi_{I1}\circ\chi_{I2}\cdots\circ\chi_{IN}$ and $\chi_{Ik}$ denotes the ideal input-output subprocesses connecting the $k$th node and the $(k+1)$th node. In the $N$-local model, the whole process $\chi_{GC1|N}\equiv\prod_{k=1}^{N}\chi_{GCk}$ is composed of the subprocesses $\chi_{GCk}$ between the $k$th node and the $(k+1)$th node, each having its own underlying LHV $\lambda_{k}$. Non-$N$locality then exists if $F_{\text{expt}1|N}\equiv\text{tr}(\chi_{I1N}\prod_{k=1}^{N}\chi_{\text{expt}k})>F_{GC1|N}\equiv \max_{\chi_{GC1|N}}\hspace{3pt}\text{tr}(\chi_{GC1|N}\chi_{I1N})$, where $F_{GC1|3}\simeq0.6768$, $F_{GC1|4}\simeq0.6250$, and $\lim_{N\rightarrow \infty}F_{GC1|N}\simeq0.5000$ (See Appendix \ref{derived_fidelities}).

Finally, we extend the idea of $F_{\text{expt112}}$ and introduce the following experimental process fidelity $F_{\text{expt}11N}\equiv\text{tr}(\chi_{I1N}\chi_{\text{expt}1}\circ\prod_{k=2}^{N}\chi_{\text{expt}1k})$, where $\chi_{\text{expt}1k}$ describes experimental teleportation from the first node in the network to the $(k+1)$th node through all $k-1$ intermediate nodes between them. Since the maximum value of $F_{\text{expt}11N}$ predicted by the $N$-local model is $F_{GC1|N}$, it can be inferred that $F_{\text{expt}11N}>F_{GC1|N}$ implies the existence of non-$N$locality of experimental teleportation.

\section{Summary and outlook}
We have proposed a formalism referred to as a genuinely classical process to characterize and identify both true quantum teleportation and the underlying quantum correlations in many-node networking teleportation. We show that quantum-information processing can be employed to quantitatively discriminate quantum correlations. The proposed formalism is well suited to the analysis of existing experiments, and faithfully evaluates the performance of all the operations required for teleportation through quantum networks.

Such a task-oriented approach raises several interesting questions, including how one can identify generic truly quantum networking tasks such as one-way quantum computation in many-node networks and what multipartite quantum correlations exist behind multipartite distributed quantum-information processing.
\\

\begin{acknowledgments}
We thank J.-W. Pan for helpful comments and discussions.  This work is partially supported by the Ministry of Science and Technology, Taiwan, under Grant Numbers MOST 107-2628-M-006-001-MY4 and MOST 107-2627-E-006-001. H.Lu was partially supported by Major Program of Shandong Province Natural Science Foundation (grants ZR2018ZB0649).
\end{acknowledgments}

\appendix

\section{IDEAL PROCESS MATRIX OF TELEPORTATION}
\label{derived_process_matrix}
The essence of PT is that a process of interest can be completely characterized by a process matrix \cite{Nielsen&Chuang00}. This process matrix is constructed by the probabilities of specific output states conditioned on specific input states $P(v^{(b)}_{j}|v^{(a)}_{i})$ as determined from the measurement data of an experimental process, where $v^{(a)}_{i}, v^{(b)}_{j}\in\{+1,-1\}$ are the $i$th and $j$th measurement results for physical properties ${V}^{(a)}_{i}$ and ${V}^{(b)}_{j}$ of the input and output respectively. The output state can be represented in the following decomposition form:
\begin{equation}
\rho_{v^{(a)}_{i}}=\frac{1}{2}(\hat{I}+\sum_{j=1}^{3}\sum_{v^{(b)}_{j}=\pm 1}v^{(b)}_{j}P(v^{(b)}_{j}|v^{(a)}_{i})\hat{V}^{(b)}_{j}),\label{state_tomographyA}
\end{equation}
where $\hat{I}$ denotes the identity matrix and $\hat{V}^{(b)}_{j}$ are three complementary observables that are the quantum analogs of the physical properties, ${V}^{(b)}_{j}$. The process matrix for the experiment can then be represented in the form
\begin{equation}
\chi_{\text{expt}} =  \frac{1}{4}\left[ \begin{matrix}
    \hat{I}_{f}+\hat{V}^{(a)}_{f,3} & \hat{V}^{(a)}_{f,1}+i\hat{V}^{(a)}_{f,2} \\
    \hat{V}^{(a)}_{f,1}-i\hat{V}^{(a)}_{f,2} & \hat{I}_{f}-\hat{V}^{(a)}_{f,3}
    \end{matrix}
\right],\label{process_matrix}
\end{equation}
where $\hat{I}_{f}\equiv\rho_{v^{(a)}_{i}=+1}+\rho_{v^{(a)}_{i}=-1}$, and $\hat{V}^{(a)}_{f,i}\equiv\rho_{v^{(a)}_{i}=+1}-\rho_{v^{(a)}_{i}=-1}$ denotes the final output matrix of observables $\hat{V}^{(a)}_{i}$ for $i=1,2,3$.

If the conditional probabilities satisfy the classical assumption
\begin{equation}
P(v^{(b)}_{j}|v^{(a)}_{i})=\sum_{\xi,\mu}P(\textbf{v}^{(a)}_{\xi}|v^{(a)}_{i})\Omega_{\xi\mu}P(v^{(b)}_{j}|\textbf{v}^{(b)}_{\mu}),\label{resultingstateA}
\end{equation}
then the process matrices of genuinely classical processes $\chi_{GC}$ can be constructed through Eqs. (\ref{state_tomographyA}) and (\ref{process_matrix}).

If the process is an ideal teleportation process \cite{Bennett96}, we get the measurement outcomes shown in Table~\ref{IdealTeleportation}. The observables used for the input states are $\hat{V}^{(a)}_{1}=X,$ $\hat{V}^{(a)}_{2}=Y,$ $\hat{V}^{(a)}_{3}=Z$. Hence, the input states for the PT algorithm are the eigenstates of the Pauli matrices, $\left|0\right\rangle,\left|1\right\rangle,\left|+\right\rangle,\left|-\right\rangle,\left|R\right\rangle,\left|L\right\rangle$, where $\left|\pm\right\rangle=(\left|0\right\rangle\pm\left|1\right\rangle)/\sqrt{2}$ and $\left|^{R}_{L}\right\rangle=(\left|0\right\rangle\pm i\left|1\right\rangle)/\sqrt{2}$.

\begin{table*}[ht]
\centering
\caption{Measurement results for ideal teleportation. The results show the case where the receiver uses the observables $\hat{V}^{(b)}_{1}=UXU^{\dag},$ $\hat{V}^{(b)}_{2}=UYU^{\dag},$ $\hat{V}^{(b)}_{3}=UZU^{\dag}$ to measure the output states of ideal teleportation, where $U=\left|0\right\rangle\!\!\left\langle0\right|+\exp(i\pi/4)\left|1\right\rangle\!\!\left\langle1\right|$. }
\begin{tabular}{|l|c|c|c|c|c|c|}
\hline
Input state &  $P({v}^{(b)}_{1}=+1|{v}^{(a)}_{i})$ & $P({v}^{(b)}_{1}=-1|{v}^{(a)}_{i})$ & $P({v}^{(b)}_{2}=+1|{v}^{(a)}_{i})$ & $P({v}^{(b)}_{2}=-1|{v}^{(a)}_{i})$ & $P({v}^{(b)}_{3}=+1|{v}^{(a)}_{i})$ & $P({v}^{(b)}_{3}=-1|{v}^{(a)}_{i})$ \\ \hline
 $\left|+\right\rangle, {v}^{(a)}_{1}=+1$ &  $(1+1/\sqrt{2})/2$ & $(1-1/\sqrt{2})/2$ & $(1-1/\sqrt{2})/2$ & $(1+1/\sqrt{2})/2$ & $1/2$ & $1/2$ \\ \hline
 $\left|-\right\rangle, {v}^{(a)}_{1}=-1$ &  $(1-1/\sqrt{2})/2$ & $(1+1/\sqrt{2})/2$ & $(1+1/\sqrt{2})/2$ & $(1-1/\sqrt{2})/2$ & $1/2$ & $1/2$ \\ \hline
 $\left|R\right\rangle, {v}^{(a)}_{2}=+1$ &  $(1+1/\sqrt{2})/2$ & $(1-1/\sqrt{2})/2$ & $(1+1/\sqrt{2})/2$ & $(1-1/\sqrt{2})/2$ & $1/2$ & $1/2$ \\ \hline
 $\left|L\right\rangle, {v}^{(a)}_{2}=-1$ &  $(1-1/\sqrt{2})/2$ & $(1+1/\sqrt{2})/2$ & $(1-1/\sqrt{2})/2$ & $(1+1/\sqrt{2})/2$ & $1/2$ & $1/2$ \\ \hline
 $\left|0\right\rangle, {v}^{(a)}_{3}=+1$ &  $1/2$ & $1/2$ & $1/2$ & $1/2$ & $1$ & $0$ \\ \hline
 $\left|1\right\rangle, {v}^{(a)}_{3}=-1$ &  $1/2$ & $1/2$ & $1/2$ & $1/2$ & $0$ & $1$ \\ \hline
\end{tabular}
\label{IdealTeleportation}
\end{table*}

The observables used for the output states are $\hat{V}^{(b)}_{1}=UXU^{\dag},$ $\hat{V}^{(b)}_{2}=UYU^{\dag},$ $\hat{V}^{(b)}_{3}=UZU^{\dag}$, where $U=\left|0\right\rangle\!\!\left\langle0\right|+\exp(i\pi/4)\left|1\right\rangle\!\!\left\langle1\right|$.
\begin{equation}
\hat{V}^{(b)}_{1}= \left[ \begin{matrix}
    0 & \frac{1-i}{\sqrt{2}} \\
    \frac{1+i}{\sqrt{2}} & 0
    \end{matrix}
\right],
\hat{V}^{(b)}_{2}= \left[ \begin{matrix}
    0 & \frac{-1-i}{\sqrt{2}} \\
    \frac{-1+i}{\sqrt{2}} & 0
    \end{matrix}
\right],
\hat{V}^{(b)}_{3}= \left[ \begin{matrix}
    1 & 0 \\
    0 & -1
    \end{matrix}
\right].\nonumber
\end{equation}

Using state tomography [Eq.~(\ref{state_tomographyA})] yields the density matrix of the output states, in which
\begin{equation}
\begin{split}
\rho_{v^{(a)}_{1}=+1}&=\frac{1}{2}(\hat{I}+\frac{1}{\sqrt{2}}\hat{V}^{(b)}_{1}-\frac{1}{\sqrt{2}}\hat{V}^{(b)}_{2}+0\hat{V}^{(b)}_{3})\\
&= \frac{1}{2}\left[ \begin{matrix}
    1 & 1 \\
    1 & 1
    \end{matrix}
\right]=\left|+\right\rangle\!\!\left\langle +\right| ,\nonumber
\end{split}
\end{equation}
\begin{equation}
\begin{split}
\rho_{v^{(a)}_{1}=-1}&=\frac{1}{2}(\hat{I}-\frac{1}{\sqrt{2}}\hat{V}^{(b)}_{1}+\frac{1}{\sqrt{2}}\hat{V}^{(b)}_{2}+0\hat{V}^{(b)}_{3})\\
&= \frac{1}{2}\left[ \begin{matrix}
    1 & -1 \\
    -1 & 1
    \end{matrix}
\right]=\left|-\right\rangle\!\!\left\langle -\right| ,\nonumber
\end{split}
\end{equation}
\begin{equation}
\begin{split}
\rho_{v^{(a)}_{2}=+1}&=\frac{1}{2}(\hat{I}+\frac{1}{\sqrt{2}}\hat{V}^{(b)}_{1}+\frac{1}{\sqrt{2}}\hat{V}^{(b)}_{2}+0\hat{V}^{(b)}_{3})\\
&= \frac{1}{2}\left[ \begin{matrix}
    1 & -i \\
    i & 1
    \end{matrix}
\right]=\left|R\right\rangle\!\!\left\langle R\right| ,\nonumber
\end{split}
\end{equation}
\begin{equation}
\begin{split}
\rho_{v^{(a)}_{2}=-1}&=\frac{1}{2}(\hat{I}-\frac{1}{\sqrt{2}}\hat{V}^{(b)}_{1}-\frac{1}{\sqrt{2}}\hat{V}^{(b)}_{2}+0\hat{V}^{(b)}_{3})\\
&= \frac{1}{2}\left[ \begin{matrix}
    1 & i \\
    -i & 1
    \end{matrix}
\right]=\left|L\right\rangle\!\!\left\langle L\right| ,\nonumber
\end{split}
\end{equation}
\begin{equation}
\begin{split}
\rho_{v^{(a)}_{3}=+1}&=\frac{1}{2}(\hat{I}+0\hat{V}^{(b)}_{1}+0\hat{V}^{(b)}_{2}+\hat{V}^{(b)}_{3})\\
&= \left[ \begin{matrix}
    1 & 0 \\
    0 & 0
    \end{matrix}
\right]=\left|0\right\rangle\!\!\left\langle 0\right| ,\nonumber
\end{split}
\end{equation}
\begin{equation}
\begin{split}
\rho_{v^{(a)}_{3}=-1}&=\frac{1}{2}(\hat{I}+0\hat{V}^{(b)}_{1}+0\hat{V}^{(b)}_{2}-\hat{V}^{(b)}_{3})\\
&= \left[ \begin{matrix}
    0 & 0 \\
    0 & 1
    \end{matrix}
\right]=\left|1\right\rangle\!\!\left\langle 1\right| .\nonumber
\end{split}
\end{equation}
The process matrix of ideal teleportation $\chi_{I}$ can then be expressed as
\begin{equation}
\begin{split}
\chi_{I}&=  \frac{1}{4}\left[ \begin{matrix}
    \hat{I}_{f}+\hat{V}_{f,3} & \hat{V}_{f,1}+i\hat{V}_{f,2} \\
    \hat{V}_{f,1}-i\hat{V}_{f,2} & \hat{I}_{f}-\hat{V}_{f,3}
    \end{matrix}\right]\\
    &= \frac{1}{2}\left[ \begin{matrix}
    1 & 0 & 0 & 1\\
    0 & 0 & 0 & 0\\
    0 & 0 & 0 & 0\\
    1 & 0 & 0 & 1
    \end{matrix}
\right].
\end{split}\label{ideal_process}
\end{equation}

In practical teleportation experiments, to objectively check teleportation between two parties (e.g., Alice and Bob), it is necessary to use a verifier to collect the measurement outcomes and build the process matrix. The verifier first sends Alice one of the eigenstates of $\{\hat{V}_{j}^{(a)}\}$ randomly, and then asks her to send it to Bob. Since the states are randomly chosen by the verifier, the input states are unknown to Alice and Bob. Bob measures the observables $\hat{V}^{(b)}_{j}$ chosen by the verifier for the received state, and then sends the measurement outcomes back to the verifier. After many runs of experiments and collecting all the data, the verifier constructs a process matrix using the PT algorithm and checks whether or not the teleportation is not a GCP using $F_{\text{expt}}\equiv \text{tr}(\chi_{\text{expt}}\chi_{I})> F_{GC}\equiv\max_{\chi_{GC}} \text{tr}(\chi_{GC}\chi_{I})$ [see Eq.~(4) in the main text].\\

\section{COMPARISON BETWEEN $\chi_{GC}$ AND $\chi_{C}$}
\label{comparison}
The relationship between input and output states of a CP, $\chi_{C}$, can be characterized by $P(v^{(b)}_{j}|v^{(a)}_{i})=\sum_{\xi,\mu}P(\textbf{v}^{(a)}_{\xi}|v^{(a)}_{i})\Omega_{\xi\mu}P(v^{(b)}_{j}|\mu)$, where $\mu$ denotes an output state that can be reconstructed as a density operator $\rho^{(b)}_{\mu}$ by using state tomography \cite{Hsieh17}. The average value of observables required in the state tomography can be derived from the conditional probability $P(v^{(b)}_{j}|\mu)$ by $\text{tr}(\hat{V}^{(b)}_{j}\rho^{(b)}_{\mu})=\sum_{v^{(b)}_{j}=\pm 1}v^{(b)}_{j}P(v^{(b)}_{j}|\mu)$.
Here one can use the classical states, $\textbf{v}^{(b)}_{\kappa}$, as used in a GCP, to represent $P(v^{(b)}_{j}|\mu)$ as $P(v^{(b)}_{j}|\mu)=\sum_{\kappa}P_{\mu\kappa}P(v^{(b)}_{j}|\textbf{v}^{(b)}_{\kappa})$, where $P_{\mu\kappa}$ is a probability distribution of classical states $\textbf{v}^{(b)}_{\kappa}$. Therefore, the $P(v^{(b)}_{j}|v^{(a)}_{i})$ for the CP can be re-expressed as $P(v^{(b)}_{j}|v^{(a)}_{i})$ $=\sum_{\xi,\mu,\kappa}P(\textbf{v}^{(a)}_{\xi}|v^{(a)}_{i})\Omega_{\xi\mu}P_{\mu\kappa}P(v^{(b)}_{j}|\textbf{v}^{(b)}_{\kappa})$. Through the transition probability $\Omega_{\xi\kappa}=\sum_{\mu}\Omega_{\xi\mu}P_{\mu\kappa}$, $P(v^{(b)}_{j}|v^{(a)}_{i})$ of the CP can be rephrased as $P(v^{(b)}_{j}|v^{(a)}_{i})=\sum_{\xi,\kappa}P(\textbf{v}^{(a)}_{\xi}|v^{(a)}_{i})\Omega_{\xi\kappa}P(v^{(b)}_{j}|\textbf{v}^{(b)}_{\kappa})$, which is in the same form as Eq.~(\ref{resultingstateA}). With Eqs.~(2) and (3) in the main text, the $\chi_{C}$ can then be described by a $\chi_{GC}$.

\section{COMPUTATIONAL COST OF THE MATHEMATICAL MAXIMIZATION TASK}
\label{Computational_cost}

For a $d$-dimensional system, the number of possible $\Omega_{\xi\mu}$ under examination is $d^{2(d^2-1)}$ and determined by the number of observables used for PT. Then the required computational cost to determine the best simulation of teleportation by $\chi_{GC}$ increases with $d$. For a given network, the dimension of the process is the only factor that dominates the computational cost.

The other factors of a process, e.g., the number of nodes $N$ in $N$-node teleportation network, does not contribute to the computational cost. The reason is that the $N$-node criteria are derived from the best mimicry $\chi_{GC}$ for $N=2$. To make the whole process in the $N$-local model, $\chi_{GC1|N}\equiv\prod_{k=1}^{N}\chi_{GCk}=\chi_{GC1}\circ\chi_{GC2}\cdots\circ\chi_{GCN}$, be the best mimicry of teleportation, it is necessary that each $\chi_{GCk}$ between the $k$th node and the $(k+1)$th node is the best mimicry of $\chi_{I}$. Thus the computational cost for the $N$-node criteria is the same as the two-node case.

\section{CLEAREST DISTINCTION BETWEEN $\chi_{I}$ AND $\chi_{GC}$}
\label{Clearest_distinction}
\begin{figure}[t]
\includegraphics[width=8.5 cm]{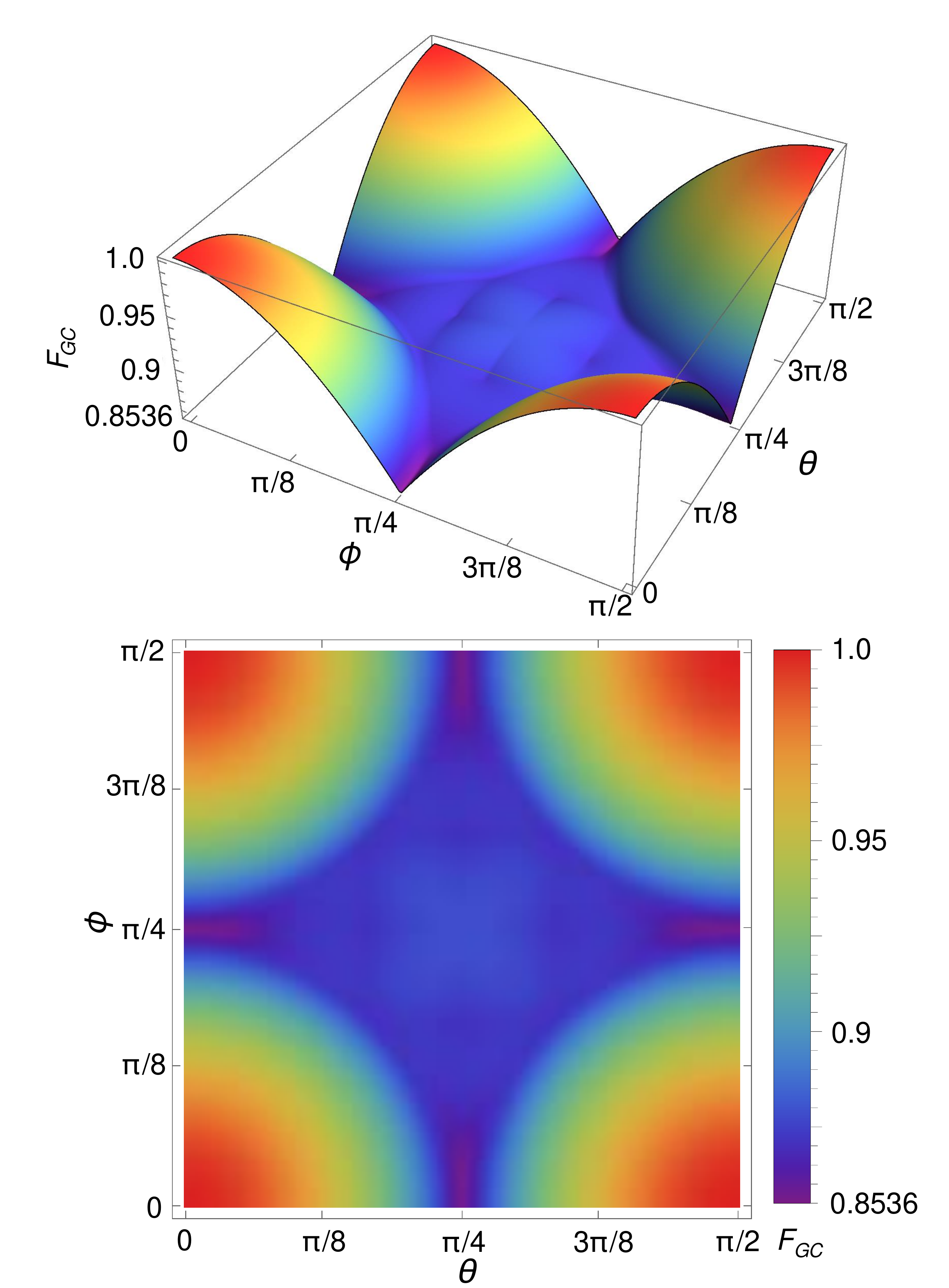}
\caption{Relationship between $F_{GC}$ and different choices of observables. Note that $F_{GC}$ is scanned through $\theta$ and $\phi$ in the unitary transform $U$ in Eq.~(\ref{unitary}) that rotates the observables.}\label{fidelity_density}
\end{figure}

Genuinely classical processes $\chi_{GC}$ have different capability to mimic ideal teleportation given different choices of the observables $\hat{V}^{(b)}_{j}$ for PT, i.e.,
\begin{equation}
F_{GC}\equiv \max_{\chi_{GC}}\hspace{3pt}\text{tr}(\chi_{GC}\chi_{I}).
\end{equation}
To find the clearest distinction possible between the quantum mechanical prediction $\chi_{I}$ and the classical result $\chi_{GC}$, we process all the complementary observables ($\hat{V}^{(b)}_{1}=UXU^{\dag},$ $\hat{V}^{(b)}_{2}=UYU^{\dag},$ $\hat{V}^{(b)}_{3}=UZU^{\dag}$) using an arbitrary unitary transform \cite{Nielsen&Chuang00}, i.e.,
\begin{equation}
U=\left[ \begin{matrix}
    e^{-i\phi/2}\cos(\frac{\theta}{2}) & e^{-i\phi/2}\sin(\frac{\theta}{2}) \\
    -e^{i\phi/2}\sin(\frac{\theta}{2}) & e^{i\phi/2}\cos(\frac{\theta}{2})
    \end{matrix}
\right].\label{unitary}
\end{equation}
The corresponding results for $F_{GC}$ are shown in Fig.~\ref{fidelity_density}. It is noted that there exist more than one set of observables that maximize the distinction between the genuinely classical process and the teleportation process. In the main text, we arbitrarily choose $\theta=0$, $\phi=\pi/4$, and $F_{GC}=0.8536$.

\begin{table*}[ht]
\centering
\caption{Measurement results for the best mimicry of ideal teleportation by $\chi_{GC}$. With the optimal set of joint probabilities $\{\Omega'_{\xi\mu}\}$ [Eq.~(\ref{trans8536})] found by SDP for the maximization task, we get the conditional probabilities $P(v^{(b)}_{j}|v^{(a)}_{i})$ [Eq.~(\ref{resultingstateA})] using the same way as illustrated by Eq.~(\ref{resultingstate_3}). The process matrix of $\chi_{GC}$ (\ref{processmatrix8536}) is obtained by substituting these conditional probabilities into Eqs.~(\ref{state_tomographyA}) and (\ref{process_matrix}).}
\begin{tabular}{|l|c|c|c|c|c|c|}
\hline
Input state &  $P({v}^{(b)}_{1}=+1|{v}^{(a)}_{i})$ & $P({v}^{(b)}_{1}=-1|{v}^{(a)}_{i})$ & $P({v}^{(b)}_{2}=+1|{v}^{(a)}_{i})$ & $P({v}^{(b)}_{2}=-1|{v}^{(a)}_{i})$ & $P({v}^{(b)}_{3}=+1|{v}^{(a)}_{i})$ & $P({v}^{(b)}_{3}=-1|{v}^{(a)}_{i})$ \\ \hline
 ${v}^{(a)}_{1}=+1$ &  $0.7500$ & $0.2500$ & $0.2500$ & $0.7500$ & $0.5000$ & $0.5000$ \\ \hline
 ${v}^{(a)}_{1}=-1$ &  $0.2500$ & $0.7500$ & $0.7500$ & $0.2500$ & $0.5000$ & $0.5000$ \\ \hline
 ${v}^{(a)}_{2}=+1$ &  $0.7500$ & $0.2500$ & $0.7500$ & $0.2500$ & $0.5000$ & $0.5000$ \\ \hline
 ${v}^{(a)}_{2}=-1$ &  $0.2500$ & $0.7500$ & $0.2500$ & $0.7500$ & $0.5000$ & $0.5000$ \\ \hline
 ${v}^{(a)}_{3}=+1$ &  $0.5000$ & $0.5000$ & $0.5000$ & $0.5000$ & $1.0000$ & $0.0000$ \\ \hline
 ${v}^{(a)}_{3}=-1$ &  $0.5000$ & $0.5000$ & $0.5000$ & $0.5000$ & $0.0000$ & $1.0000$ \\ \hline
\end{tabular}
\label{tabletwo}
\end{table*}

We proceed to show how the best $\chi_{GC}$ works to mimic an ideal teleportation to attain the best process fidelity, $F_{GC}=0.8536$. Any $\chi_{GC}$ is described by the classical states and its evolution. The classical states  $\textbf{v}_{\xi}(\text{v}_{1},\text{v}_{2},\text{v}_{3})$ of input and output systems satisfying the assumption of realism can be concretely represented by the following sets of physical properties $V_{1}$, $V_{2}$, and $V_{3}$:
\begin{equation}
\begin{split}
\textbf{v}_{1}(+1,+1,+1)&,\hspace{1pt}\textbf{v}_{2}(+1,+1,-1),\\
\textbf{v}_{3}(+1,-1,+1)&,\hspace{1pt}\textbf{v}_{4}(+1,-1,-1),\\
\textbf{v}_{5}(-1,+1,+1)&,\hspace{1pt}\textbf{v}_{6}(-1,+1,-1),\\
\textbf{v}_{7}(-1,-1,+1)&,\hspace{1pt}\textbf{v}_{8}(-1,-1,-1).
\end{split}\label{setintorho}
\end{equation}
Then these classical states evolve from $\textbf{v}^{(a)}_{\xi}$ to $\textbf{v}^{(b)}_{\mu}$ through transition probabilities $\Omega_{\xi\mu}$. Their relationship is characterized by Eq.~(\ref{resultingstateA}).

Next, to perform process tomography on GCP, we consider specific states of physical properties as the input states, e.g., $v^{(a)}_{1}=+1$, and their subsequent time evolution.

According to Eq.~(\ref{state_tomographyA}), the output states $\rho_{v^{(a)}_{1}=+1}$ can be expressed by the conditional probabilities $P(v^{(b)}_{j}|v^{(a)}_{i})$ as shown in Eq.~(\ref{resultingstateA}). The $P(\textbf{v}^{(a)}_{\xi}|v^{(a)}_{i})$ in Eq.~(\ref{resultingstateA}) can be rephrased as $2P(\textbf{v}^{(a)}_{\xi})P(v^{(a)}_{i}|\textbf{v}^{(a)}_{\xi})$, where the assumption of an uniform probability distribution of $v^{(a)}_{i}$, i.e., $P(v^{(a)}_{i}=+1)=P(v^{(a)}_{i}=-1)=1/2$ $\forall i$, and the relation for the joint
probability $P(v^{(a)}_{i})P(\textbf{v}^{(a)}_{\xi}|v^{(a)}_{i})=P(\textbf{v}^{(a)}_{\xi})P(v^{(a)}_{i}|\textbf{v}^{(a)}_{\xi})$ have been used.
Then Eq.~(\ref{resultingstateA}) becomes
\begin{equation}
P(v^{(b)}_{j}|v^{(a)}_{i})=\sum_{\xi,\mu}2P(\textbf{v}^{(a)}_{\xi})P(v^{(a)}_{i}|\textbf{v}^{(a)}_{\xi})\Omega_{\xi\mu}P(v^{(b)}_{j}|\textbf{v}^{(b)}_{\mu}).\label{resultingstate_2}
\end{equation}
The conditional probabilities $P(v^{(a)}_{i}|\textbf{v}^{(a)}_{\xi})$ and $P(v^{(b)}_{j}|\textbf{v}^{(b)}_{\mu})$ in Eq.~(\ref{resultingstate_2}) are deterministic according to Eq.~(\ref{setintorho}). Let us take $P(v^{(b)}_{1}=+1|v^{(a)}_{1}=+1)$ for example. Since we have $P(v_{1}=+1|\textbf{v}_{\xi})=1,\ P(v_{1}=-1|\textbf{v}_{\xi})=0$ for $\xi\in\{1,2,3,4\}$ and $P(v_{1}=+1|\textbf{v}_{\xi})=0,\ P(v_{1}=-1|\textbf{v}_{\xi})=1$ for $\xi\in\{5,6,7,8\}$, the conditional probability can be expressed as
\begin{equation}
\begin{split}
P(v^{(b)}_{1}=+1|v^{(a)}_{1}=+1)&=\sum_{\xi=1}^4\sum_{\mu=1}^4 2P(\textbf{v}^{(a)}_{\xi})\Omega_{\xi\mu} \\
&=\sum_{\xi=1}^4\sum_{\mu=1}^4 \Omega'_{\xi\mu}.
\end{split}\label{resultingstate_3}
\end{equation}
The $\Omega'_{\xi\mu}=2P(\textbf{v}^{(a)}_{\xi})\Omega_{\xi\mu}$ are the joint probabilities under evaluation for performing the maximization task via SDP [see Eq.~(4) and its explanation]. Using Eq.~(\ref{state_tomographyA}), we get the density matrix of $\rho_{v^{(a)}_{1}=+1}$. The above derivation of $\rho_{v^{(a)}_{1}=+1}$ can be applied to the states $\rho_{v^{(a)}_{1}=-1}$, $\rho_{v^{(a)}_{2}=\pm1}$, and $\rho_{v^{(a)}_{3}=\pm1}$. Therefore, by Eq.~(\ref{process_matrix}), the process matrix of $\chi_{GC}$ can be explicitly expressed by the output states, $\rho_{v^{(a)}_{1}=\pm1}$, $\rho_{v^{(a)}_{2}=\pm1}$, and $\rho_{v^{(a)}_{3}=\pm1}$, i.e., $\chi_{GC}$ can be described by the corresponding joint probabilities $\Omega'_{\xi\mu}$ under examination.

With the structure of $\chi_{GC}$ shown above, we proceed to describe how the process matrix of the best teleportation simulation with $F_{GC}=0.8536$ can be concretely constructed via $\Omega'_{\xi\mu}$. The optimal set of joint probabilities $\{\Omega'_{\xi\mu}\}$ found by SDP for the maximization task can be represented in the following matrix form:
 \begin{widetext}
 \begin{equation}
 \Omega'=\left[ \begin{matrix}
    0.125 & 0 & 0.125 & 0 & 0 & 0 & 0 & 0\\
    0 & 0.125 & 0 & 0.125 & 0 & 0 & 0 & 0\\
    0 & 0 & 0.125 & 0 & 0 & 0 & 0.125 & 0\\
    0 & 0 & 0 & 0.125 & 0 & 0 & 0 & 0.125\\
    0.125 & 0 & 0 & 0 & 0.125 & 0 & 0 & 0\\
    0 & 0.125 & 0 & 0 & 0 & 0.125 & 0 & 0\\
    0 & 0 & 0 & 0 & 0.125 & 0 & 0.125 & 0\\
    0 & 0 & 0 & 0 & 0 & 0.125 & 0 & 0.125
    \end{matrix}
\right],\label{trans8536}
\end{equation}
\end{widetext}
where the matrix element of the $\xi$th column and the $\mu$th row represents the joint probability $\Omega'_{\xi\mu}$. For example, the joint probability of $\textbf{v}^{(a)}_{1}$ and $\textbf{v}^{(b)}_{1}$ is $\Omega'_{11}=0.125$, which can be read from Eq.~(\ref{trans8536}).

With the full knowledge about the optimal set of joint probabilities $\{\Omega'_{\xi\mu}\}$ [Eq.~(\ref{trans8536})], the conditional probabilities $P(v^{(b)}_{j}|v^{(a)}_{i})$ [Eq.~(\ref{resultingstateA})] can then be determined. For example, we have $P(v^{(b)}_{1}=+1|v^{(a)}_{1}=+1)=0.7500$ by using Eqs.~(\ref{resultingstate_3}) and (\ref{trans8536}). All the other conditional probabilities are listed in Table II. Using these measurement results and Eq.~(\ref{state_tomographyA}), we can get the density matrices of output states $\rho_{v^{(a)}_{i}=\pm1}$; by which, together with Eq.~(\ref{process_matrix}), the process matrix of $\chi_{GC}$ for the best simulation of teleportation with $F_{GC}=0.8536$ is as follows
\begin{equation}
\chi_{GC}= \frac{1}{2}\left[ \begin{matrix}
    1 & 0 & 0 & \frac{1}{\sqrt{2}}\\
    0 & 0 & 0 & 0\\
    0 & 0 & 0 & 0\\
    \frac{1}{\sqrt{2}} & 0 & 0 & 1
    \end{matrix}
\right].\label{processmatrix8536}
\end{equation}
It is noted that the distribution $P(\textbf{v}^{(a)}_{\xi})$ and transition probabilities $\Omega_{\xi\mu}$ can be derived from the set of joint probabilities $\{\Omega'_{\xi\mu}\}$. For $\Omega'_{\xi\mu}=2P(\textbf{v}^{(a)}_{\xi})\Omega_{\xi\mu}$ and $\sum_{\mu}\Omega_{\xi\mu}=1$, the distribution $P(\textbf{v}^{(a)}_{\xi})$ under a given $\xi$ can be derived as follows: $\sum_{\mu}\Omega'_{\xi\mu}=\sum_{\mu}2P(\textbf{v}^{(a)}_{\xi})\Omega_{\xi\mu}=2P(\textbf{v}^{(a)}_{\xi})\sum_{\mu}\Omega_{\xi\mu}=2P(\textbf{v}^{(a)}_{\xi})$, implying that $P(\textbf{v}^{(a)}_{\xi})=1/2\sum_{\mu}\Omega'_{\xi\mu}$. For the the optimal set of joint probabilities shown in Eq.~(\ref{trans8536}), we have $P(\textbf{v}^{(a)}_{\xi})=1/8$.

Equation~(\ref{processmatrix8536}) means that, after the dynamical process $\chi_{GC}$, the input state $\rho$ becomes
\begin{equation}
\chi_{GC}(\rho) = 0.8536\hat{I}\rho\hat{I}^{\dag}+0.1464Z\rho Z^{\dag}.
\end{equation}
The effect of the process $\chi_{GC}$ in the operator-sum representation as shown above describes that the input state $\rho$ undergoes a phase damping channel with noise intensity $0.1464$ \cite{Nielsen&Chuang00}. Thus the best simulation of ideal teleportation works as a phase damping process.

\section{PROCESS LOCAL HIDDEN VARIABLES MODEL}
\label{process_LHV}
A GCP can be treated as an input-output transformation implemented by sharing LHVs between the parties involved (Alice and Bob) for a networking process. To explicitly see the role played by the LHVs, let Eq.~(\ref{resultingstate}) be rephrased as $P(v^{(b)}_{j}|v^{(a)}_{i})=2\sum_{\xi,\mu}P(v^{(a)}_{i}|\textbf{v}^{(a)}_{\xi})P(\textbf{v}^{(a)}_{\xi},\textbf{v}^{(b)}_{\mu})P(v^{(b)}_{j}|\textbf{v}^{(b)}_{\mu})$, where $P(\textbf{v}^{(a)}_{\xi},\textbf{v}^{(b)}_{\mu})=P(\textbf{v}^{(a)}_{\xi})\Omega_{\xi\mu}$ is the joint probability of $\textbf{v}^{(a)}_{\xi}$ and $\textbf{v}^{(b)}_{\mu}$. Note that the relation for the joint probability $P(\textbf{v}^{(a)}_{\xi})P(v^{(a)}_{i}|\textbf{v}^{(a)}_{\xi})=P(v^{(a)}_{i})P(\textbf{v}^{(a)}_{\xi}|v^{(a)}_{i})$ and the assumption of an equal probability of the input states $P(v^{(a)}_{i})=1/2$ have both been applied above. Furthermore, let each of the joint outcome sets $(\textbf{v}^{(a)}_{\xi},\textbf{v}^{(b)}_{\mu})$ be assigned a LHV $\lambda$. Then $P(v^{(b)}_{j}|v^{(a)}_{i})$ can be explained using LHV $\lambda$. Equation~(\ref{resultingstate}) in a so-called LHV model for input and output states thus becomes $P(v^{(b)}_{j}|v^{(a)}_{i})=2\sum_{\lambda}P(v^{(a)}_{i}|\lambda)P(\lambda)P(v^{(b)}_{j}|\lambda)$.

\section{DERIVATIONS OF $F_{GC1|2}$, $F_{GC1|C2}$, AND $F_{GC1|N}$}
\label{derived_fidelities}
To derive the fidelity threshold of the hybrid process, $F_{GC12}\equiv \max_{\chi_{GC12}}\text{tr}(\chi_{GC12}\chi_{I12})$, we first need to calculate the process matrix of a hybrid process $\chi_{GC1|2}\equiv\chi_{GC1}\circ\chi_{GC2}$. For $\chi_{GC1|2}$, the process matrix can be derived through the matrix elements $\chi_{i,n}$ in $\chi_{GCi}$ for $i\in\{1,2\}$, where
\begin{equation}
\chi_{GCi}= \frac{1}{2}\left[ \begin{matrix}
    \chi_{i,1} & \chi_{i,2} & \chi_{i,3} & \chi_{i,4}\\
    \chi_{i,5} & \chi_{i,6} & \chi_{i,7} & \chi_{i,8}\\
    \chi_{i,9} & \chi_{i,10} & \chi_{i,11} & \chi_{i,12}\\
    \chi_{i,13} & \chi_{i,14} & \chi_{i,15} & \chi_{i,16}
    \end{matrix}
\right].
\end{equation}
The hybrid process matrix can then be expressed as
\begin{equation}
\chi_{GC1|2}= \frac{1}{2}\left[ \begin{matrix}
    \chi_{12,1} & \chi_{12,2} & \chi_{12,3} & \chi_{12,4}\\
    \chi_{12,5} & \chi_{12,6} & \chi_{12,7} & \chi_{12,8}\\
    \chi_{12,9} & \chi_{12,10} & \chi_{12,11} & \chi_{12,12}\\
    \chi_{12,13} & \chi_{12,14} & \chi_{12,15} & \chi_{12,16}
    \end{matrix}\label{joint_process}
\right],
\end{equation}
in which the matrix elements are given by
\begin{equation}
\begin{split}
&\chi_{12,1}=\chi_{1,1}\chi_{2,1}+\chi_{1,2}\chi_{2,3}+\chi_{1,5}\chi_{2,9}+\chi_{1,6}\chi_{2,11},\\
&\chi_{12,2}=\chi_{1,1}\chi_{2,2}+\chi_{1,2}\chi_{2,4}+\chi_{1,5}\chi_{2,10}+\chi_{1,6}\chi_{2,12},\\
&\chi_{12,3}=\chi_{1,3}\chi_{2,1}+\chi_{1,4}\chi_{2,3}+\chi_{1,7}\chi_{2,9}+\chi_{1,8}\chi_{2,11},\\
&\chi_{12,4}=\chi_{1,3}\chi_{2,2}+\chi_{1,4}\chi_{2,4}+\chi_{1,7}\chi_{2,10}+\chi_{1,8}\chi_{2,12},\\
&\chi_{12,5}=\chi_{1,1}\chi_{2,5}+\chi_{1,2}\chi_{2,7}+\chi_{1,5}\chi_{2,13}+\chi_{1,6}\chi_{2,15},\\
&\chi_{12,6}=\chi_{1,1}\chi_{2,6}+\chi_{1,2}\chi_{2,8}+\chi_{1,5}\chi_{2,14}+\chi_{1,6}\chi_{2,16},\\
&\chi_{12,7}=\chi_{1,3}\chi_{2,5}+\chi_{1,4}\chi_{2,7}+\chi_{1,7}\chi_{2,13}+\chi_{1,8}\chi_{2,15},\\
&\chi_{12,8}=\chi_{1,3}\chi_{2,6}+\chi_{1,4}\chi_{2,8}+\chi_{1,7}\chi_{2,14}+\chi_{1,8}\chi_{2,16},\\
&\chi_{12,9}=\chi_{1,9}\chi_{2,1}+\chi_{1,10}\chi_{2,3}+\chi_{1,13}\chi_{2,9}+\chi_{1,14}\chi_{2,11},\\
&\chi_{12,10}=\chi_{1,9}\chi_{2,2}+\chi_{1,10}\chi_{2,4}+\chi_{1,13}\chi_{2,10}+\chi_{1,14}\chi_{2,12},\\
&\chi_{12,11}=\chi_{1,11}\chi_{2,1}+\chi_{1,12}\chi_{2,3}+\chi_{1,15}\chi_{2,9}+\chi_{1,16}\chi_{2,11},\\
&\chi_{12,12}=\chi_{1,11}\chi_{2,2}+\chi_{1,12}\chi_{2,4}+\chi_{1,15}\chi_{2,10}+\chi_{1,16}\chi_{2,12},\\
&\chi_{12,13}=\chi_{1,9}\chi_{2,5}+\chi_{1,10}\chi_{2,7}+\chi_{1,13}\chi_{2,13}+\chi_{1,14}\chi_{2,15},\\
&\chi_{12,14}=\chi_{1,9}\chi_{2,6}+\chi_{1,10}\chi_{2,8}+\chi_{1,13}\chi_{2,14}+\chi_{1,14}\chi_{2,16},\\
&\chi_{12,15}=\chi_{1,11}\chi_{2,5}+\chi_{1,12}\chi_{2,7}+\chi_{1,15}\chi_{2,13}+\chi_{1,16}\chi_{2,15},\\
&\chi_{12,16}=\chi_{1,11}\chi_{2,6}+\chi_{1,12}\chi_{2,8}+\chi_{1,15}\chi_{2,14}+\chi_{1,16}\chi_{2,16}.
\end{split}
\end{equation}

The process matrix $\chi_{GC}$ with the best capability of genuinely classical mimicry for $\chi_{I}$ found by SDP is in Eq.~(\ref{processmatrix8536}).
Using Eq.~(\ref{joint_process}) and $\chi_{GC2}=\chi_{GC1}=\chi_{GC}$, we obtain the process matrix
  \begin{equation}
  \begin{split}
\chi_{GC1|2}&=\chi_{GC1}\circ\chi_{GC2}\\
&= \frac{1}{2}\left[ \begin{matrix}
    1 & 0 & 0 & \frac{1}{2}\\
    0 & 0 & 0 & 0\\
    0 & 0 & 0 & 0\\
    \frac{1}{2} & 0 & 0 & 1
    \end{matrix}
\right],
\end{split}
\end{equation}
and then calculate the fidelity $F_{GC1|2}\equiv\text{tr}(\chi_{GC1|2}\chi_{I})=0.7500$.\\

\begin{figure}[t]
\includegraphics[width=7.5cm]{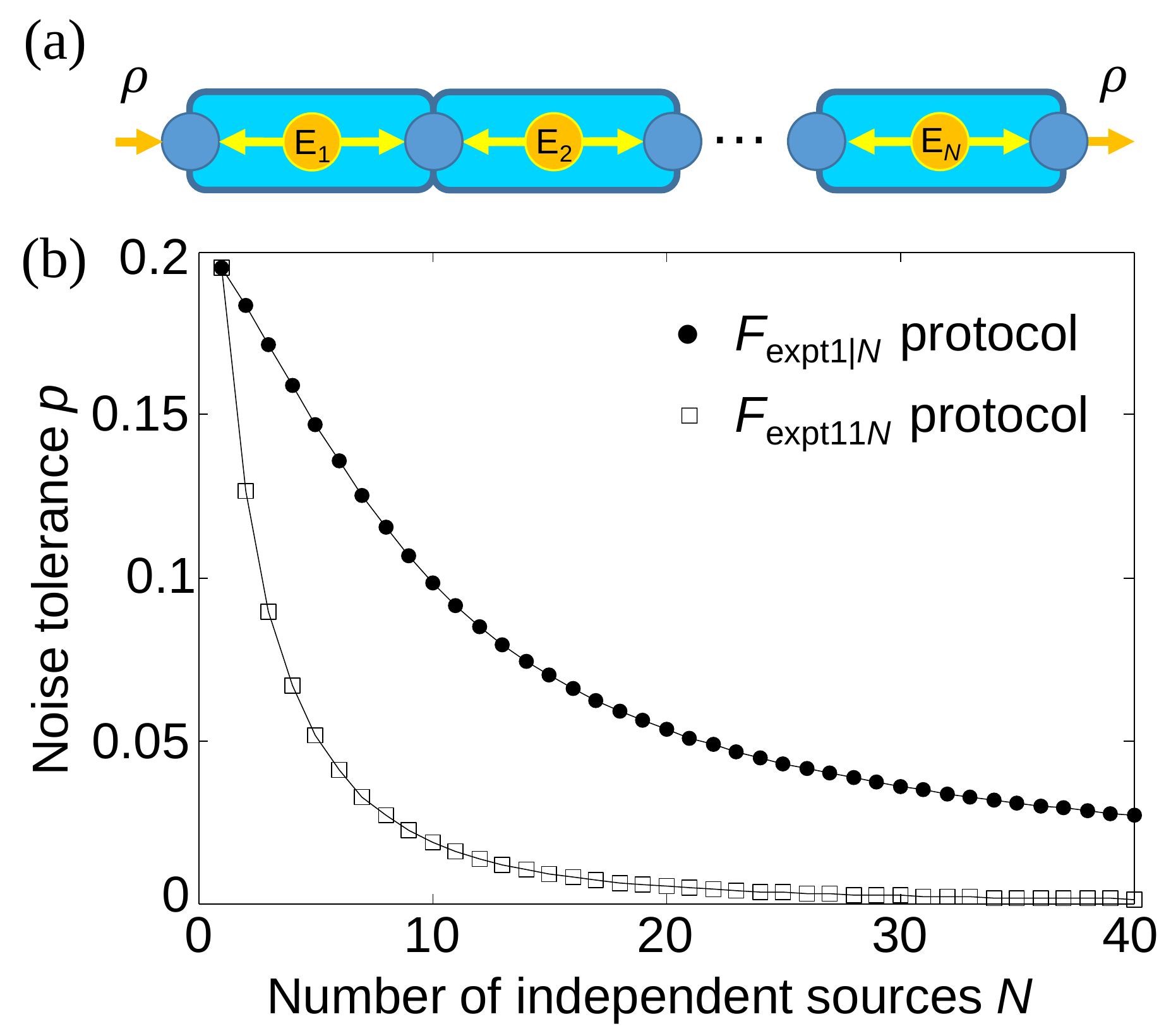}
\caption{Noise tolerance of non-$N$ locality criteria. Suppose that a minimum amount of white noise ($p$) is added to each entanglement source (a) such that the noisy experimental process cannot be identified by the proposed criteria, i.e.,  (b) $F_{\text{expt}1|N}=F_{GC1|N}$ (dots) and $F_{\text{expt}11N}=F_{GC1|N}$ (squares). The robustness of the identification protocol is degraded with increasing $N$. However, compared with the identification protocol using $F_{\text{expt}11N}$, that with $F_{\text{expt}1|N}$ is more robust against noise as $N$ increases.}\label{error_tolerance}
\end{figure}

To calculate $F_{GC1|N}\equiv\text{tr}(\chi_{GC1|N}\chi_{I})$, we use Eq.~(\ref{joint_process}) $N-1$ times. For example, $\chi_{GC1|3}=\chi_{GC1|2}\circ\chi_{GC}$. The process matrix has the form
\begin{equation}
\chi_{GC1|N}= \frac{1}{2}\left[ \begin{matrix}
    1 & 0 & 0 & (\frac{1}{\sqrt{2}})^N\\
    0 & 0 & 0 & 0\\
    0 & 0 & 0 & 0\\
    (\frac{1}{\sqrt{2}})^N & 0 & 0 & 1
    \end{matrix}
\right],
\end{equation}
and $F_{GC1|N}\equiv\text{tr}(\chi_{GC1|N}\chi_{I})=\frac{1+(\frac{1}{\sqrt{2}})^N}{2}$. Notably, the fidelity criterion $F_{\text{expt}1|N}>F_{GC1|N}$ is robust against noise even for large $N$, as shown in Fig.~\ref{error_tolerance}.\\

For the fidelity $F_{GC1|C2}\equiv\text{tr}(\chi_{GC1|C2}\chi_{I})$, the process matrix $\chi_{GC1|C2}$ can be obtained by $\chi_{C}$ and $\chi_{GC}$, where
\begin{equation}
\chi_{C}= \frac{1}{2}\left[ \begin{matrix}
    0.7887 & 0 & 0 & 0.5774\\
    0 & 0.2113 & 0 & 0\\
    0 & 0 & 0.2113 & 0\\
    0.5774 & 0 & 0 & 0.7887
    \end{matrix}
\right]
\end{equation}
is the classical process \cite{Hsieh17} that has the best classical mimicry to teleportation. The hybrid process is given by
\begin{equation}
\begin{split}
\chi_{GC1|C2}&=\chi_{GC}\circ\chi_{C}\\
&= \frac{1}{2}\left[ \begin{matrix}
    0.7887 & 0 & 0 & 0.4082\\
    0 & 0.2113 & 0 & 0\\
    0 & 0 & 0.2113 & 0\\
    0.4082 & 0 & 0 & 0.7887
    \end{matrix}
\right],
\end{split}
\end{equation}
and the fidelity $F_{GC1|C2}\equiv\text{tr}(\chi_{GC1|C2}\chi_{I})=0.5985$.


\begin{thebibliography}{10}
\bibitem{four_input}While there are six different input states in Eq.~(\ref{process_tomography}), only four different input states are needed for PT: $v^{(a)}_{1}=+1$, $v^{(a)}_{2}=+1$, $v^{(a)}_{3}=+1$, and $v^{(a)}_{3}=-1$. The reason is because $\hat{V}_{gc,i}$ can be reexpressed as $\hat{V}_{gc,i}=2\rho_{v^{(a)}_{i}=+1}-\hat{I}_{gc}$ $=2\rho_{v^{(a)}_{i}=+1}-(\rho_{v^{(a)}_{3}=+1}+\rho_{v^{(a)}_{3}=-1})$. The relation $\hat{I}_{gc}\equiv\rho_{v^{(a)}_{i}=+1}+\rho_{v^{(a)}_{i}=-1}$, for $i=1,2,3$, is used above.
\bibitem{Measure-prepare_strategy} In the measure-prepare strategy, the sender (say, Alice) directly measures an unknown state given to her by the verifier and then tells the receiver (say, Bob) the measurement result such that he can prepare the output state (see Ref.~\cite{Massar95}).
\end{thebibliography}
\end{document}